\documentclass[comsoc,journal]{IEEEtran} 

\usepackage{tablefootnote}
\usepackage{float}
\usepackage{cite}
\usepackage{graphicx}
\usepackage{newtxmath}
\usepackage{amsfonts}
\usepackage{amsmath}
\usepackage{algorithm,algcompatible}
\usepackage{fancyhdr}
\usepackage{xcolor}
\usepackage{multirow}
\usepackage{flafter}
\usepackage{url}
\usepackage{enumerate}
\usepackage{epstopdf}

\providecommand{\keywords}[1]{\textbf{\textit{Index terms---}} #1}
\newcommand{\ahmad}[1]{{\color{black}#1}}
\newcommand{\jiadong}[1]{{\color{black}#1}}

\begin{document}
\title{Attention-based QoE-aware Digital Twin Empowered Edge Computing for Immersive \\Virtual Reality}

\author{Jiadong~Yu,\IEEEmembership{ Member,~IEEE,} 
Ahmad~Alhilal,\IEEEmembership{ Member,~IEEE,}
Tailin~Zhou,\\
Pan~Hui,\IEEEmembership{ Fellow,~IEEE,}
Danny~H.K.~Tsang,\IEEEmembership{ Life Fellow,~IEEE} 
\thanks{\protect J. Yu is with the Internet of Things Thrust, The Hong Kong University of Science and Technology (Guangzhou), Guangzhou, Guangdong, China (Email: jiadongyu@ust.hk).}
\thanks{\protect A. Alhilal and T. Zhou are with IPO, Division of EMIA, The Hong Kong University of Science and Technology, Clear Water Bay, Hong Kong SAR, China (Email: aalhilal@connect.ust.hk, tzhouaq@connect.ust.hk).}
\thanks{\protect P. Hui is with the Computational Media and Arts thrust, The Hong Kong University of Science and Technology (Guangzhou), Guangzhou, Guangdong China, and also with Emerging Interdisciplinary Areas and the HKUST-DT Systems and Media Lab, The Hong Kong University of Science and Technology, Clear Water Bay, Hong Kong SAR, China (Email: panhui@ust.hk).}
\thanks{\protect D.H.K. Tsang is with the Internet of Things Thrust, The Hong Kong University of Science and Technology (Guangzhou), Guangzhou, Guangdong, China, and also with the Department of Electronic and Computer Engineering, The Hong Kong University of Science and Technology, Clear Water Bay, Hong Kong SAR, China (Email: eetsang@ust.hk).
}}

\maketitle

\begin{abstract}
Metaverse applications such as virtual reality (VR) content streaming, require optimal resource allocation strategies for mobile edge computing (MEC) to ensure a high-quality user experience. In contrast to online reinforcement learning (RL) algorithms, which can incur substantial communication overheads and longer delays, the majority of existing works employ offline-trained RL algorithms for resource allocation decisions in MEC systems. 
However, they neglect the impact of desynchronization between the physical and digital worlds on the effectiveness of the allocation strategy. In this paper, we tackle this desynchronization using a continual RL framework that facilitates the resource allocation dynamically for MEC-enabled VR content streaming. We first design a digital twin-empowered edge computing (DTEC) system and formulate a quality of experience (QoE) maximization problem based on attention-based resolution perception. This problem optimizes the allocation of computing and bandwidth resources while adapting the attention-based resolution of the VR content. The continual RL framework in DTEC enables %continual learning and 
adaptive online execution in a time-varying environment. The reward function is defined based on the QoE and horizon-fairness QoE (hfQoE) constraints. Furthermore, we propose freshness prioritized experience replay - continual deep deterministic policy gradient (FPER-CDDPG) to enhance the performance of continual learning in the presence of time-varying DT updates. We test FPER-CDDPG using extensive experiments and evaluation. FPER-CDDPG outperforms the benchmarks in terms of average latency, QoE, and successful delivery rate as well as meeting the hfQoE requirements and performance over long-term execution while ensuring system scalability with the increasing number of users.
\end{abstract}


\section{Introduction}
\label{sec:intro}

The Metaverse seeks to deliver an immersive and interactive experience to its users.  As extended reality (XR) applications such as augmented reality (AR), and virtual reality (VR) are delay-sensitive, longer end-to-end latencies lead to lower user experience due to the increasing lag between user input and its reflection on the user's display. As a result, users may experience motion sickness, which may manifest as mild discomfort, dizziness, nausea, and even vomiting\cite{xu2022full}. Metaverse systems offload computationally heavy operations (e.g., rendering) to distant powerful servers to compensate for the restricted capacity of users' devices. Therefore, rendering and downloading are the primary time-consuming tasks. Since the MEC servers are placed closer to end users, the Metaverse applications can offload the rendering tasks to Mobile edge computing (MEC) servers to reduce the download latency~\cite{9363323}. In the field of view (FoV) rendering case study~\cite{8038375}, the MEC servers take the users' events, render their FoVs, and return the scenes as video sequences to the users' devices (e.g., head-mounted displays, augmented reality glasses) to be played back.

Digital twins (DTs) are digital replicas of physical objects, processes, or systems. DTs can help to track, analyze, and optimize the physical systems~\cite{9711524}. DT-empowered edge computing (DTEC)~\cite{yu20226g} is an emerging architecture that combines MEC with DT technologies. DTEC maintains the computation state of the MEC systems (e.g., CPU clock frequency) and the communication state of the edge network (e.g., bandwidth and network state). 
Using DTEC, the MEC system state can be observed at any time. In other words, DTEC can not only collect real-world data but also fabricate emulation data.
These data can help to train the DT model to find the optimal strategy for resource allocation. Accordingly, DTEC can make decisions and send feedback to the MEC system to assign the physical resources to the users. This ensures the satisfactory quality of experience (QoE) and fair allocation of the MEC system's resources.

The human vision is hierarchical and selective. Humans pay attention to certain regions in their FoV and perceive them with higher quality than the rest. The vision hierarchy revolves primarily around three levels (i.e., central, paracentral, and peripheral). Central vision is perceived with the highest quality and covers less than $5\%$ of human vision. Paracentral vision covers around $30\%$ of human vision and allows for the perception of colors. Peripheral vision covers about $60\%$ of human vision and gives a perception of motion~\cite{fovr2019yang}. Attention-based rendering in VR streaming leverages this fact to improve the quality and efficiency of VR streaming over wireless networks. This can significantly reduce the overall computational requirements for rendering complex scenes without compromising the quality of experience (QoE). This mechanism has significant implications for the development of user-centric Metaverse applications (AR, VR, and XR applications), particularly applications that require real-time rendering of high-fidelity graphics and visual effects such as VR gaming.

\jiadong{In this paper, we employ MEC to perform VR rendering and streaming to VR users. The states of users and the MEC system are modeled using their respective DTs, which are maintained on the edge server. The user's DT captures attention information, while the MEC system's DT (DTEC) includes information such as communication throughput and computation frequency. By leveraging these DTs, attention-based rendering and resource allocation strategies are developed, with a focus on optimizing user perception as the primary QoE metric.}

\subsection{Related Works}
\label{sec:works}
%work in digital twin edge network
In recent years, offloading computationally-intensive tasks to edge servers has been widely studied. Many existing works on MEC focus primarily on designing offloading schemes and resource allocation decisions to provide a trade-off between computing latency and energy consumption of user devices based on the static environment or states known in advance~\cite{8030322}. However, the optimal allocation strategies should be designed based on the constantly perceived environmental information. This means the allocation decision should consider not only the state of the user’s surrounding environment in offloading tasks but also the time-varying user behaviour in the long run.
Since DTEC can capture the time-varying features of edge networks while providing an efficient way to train the decision-making model for system management, there are some emerging works that considered this condition and focused on the research of DT edge networks to support task-offloading decision-making~\cite{9174795,9447819,9170905}. In~\cite{9174795}, DTs of edge servers are leveraged to estimate the states of the edge servers and a DT of the entire MEC system for training an offloading scheme. The scheme utilizes a deep reinforcement learning algorithm to learn the policy that minimizes the offloading latency. The work~\cite{9447819} presents a DT-assisted scheme to cooperatively offload tasks to mobile-edge servers. The scheme is implemented through two learning algorithms, the decision tree algorithm and the double deep-Q-learning algorithm. Moreover, a digital twin wireless network  (DTWN) that incorporates DTs into wireless networks is proposed in~\cite{9170905}. DTWN contains a blockchain-empowered federated learning framework for collaborative learning. The method also utilizes multi-agent reinforcement learning that takes into account DT association, training data batch size, and bandwidth allocation to optimize edge association. This is intended to improve the system's reliability, security, and data privacy while also increasing the running efficiency and reducing the system's time cost.

%edge computing for the Metaverse
Metaverse applications require real-time processing and delivery of high-quality multimedia content to provide an interactive Metaverse experience. Owing to the resources (e.g., limited computation and battery) of the user devices (e.g., head-mounted displays), offloading of computationally-intensive tasks to MEC servers is necessary to meet these requirements. In AR remote Live support applications~\cite{7915547}, for instance, the AR Edge Computing architecture offloads the heavy algorithms to a high-end PC to improve the real-timeliness of AR display. Additionally, a proof-of-concept for remote 360-degree stereo VR gaming was presented in ~\cite{8416431}, which utilizes an edge server to run the game and perform the rendering. The server receives the user's FoV and control events obtained from the head-mounted display (HMD) and its controllers. The framework proved to save energy and computation load on the end terminals while achieving low latency and a high bit rate. The work~\cite{9789210} presents a transmission framework to minimize the average inter-player delay in multiplayer interactive VR gaming.  The framework optimizes the computing resource allocation of the MEC server, the wireless bandwidth allocation, and the post-processing decision policy subject to the constraints of the absolute delay requirements, the local energy limits of players, the total bandwidth limit, and the computing resources limit.  %The framework models the interaction among players and makes reasonable post-processing decisions  at the MEC server or mobile VR device. 
For a more general VR content delivery example, the work in~\cite{9881517} explores the advantage of offloading VR computation to edge servers to provide immersive VR services. Specifically, the work evaluates various performance metrics such as frame rate, packet loss rate, and image quality using various configuration settings. For delivering high-fidelity VR video over multi-cell MEC networks, the work in~\cite{9350227} proposes a rendering-aware tile caching scheme to optimize the end-to-end latency.

%QoE definition
To evaluate the performance of MEC-supported Metaverse in user-centric applications (VR content streaming), numerous works have defined customized QoE from different perspectives~\cite{10066518,10101693,10005856,9097455, 9565222}. The work~\cite{10066518} defines visual QoE for VR gaming. Objects in VR games (e.g., an avatar, weapon, and car) can be rendered independently.  Additionally, VR/3D games often provide multiple rendering levels, thus diverse weights and bit rates. This helps to reduce computational and network resources by using lower levels for objects of less user interest. The work~\cite{10101693} defines the QoE based on the cache hit ratio of users' requests in the edge server and VR content transmission delay. Through a collaborative architecture for background and interactive VR content generation, it distributes the caching of the VR content between edge servers according to the user's request. In~\cite{10005856}, the QoE utility is defined based on meeting tolerable motion-to-photon latency. From the general perspective of VR interactive streaming, \cite{9565222} defines the QoE as a combination of viewpoint prediction, video quality, and video jitter. 

\textit{While attention-based rendering can reduce both computation and bandwidth costs,
these works do not consider the QoE as a combination of latency and attention-based user perception. Additionally, there is an absence of works that utilize edge computing and empower it with DTs and such a QoE combination for an immersive and interactive VR experience.
}

% Since \cite{10101693} emphasized the content generation and synchronization of VR from the efficiency and low delay aspects, they defined the QoE based on whether the cache update content can satisfy the user demand at the next moment. 
% In~\cite{10005856}, the QoE utility is defined based on whether the motion-to-photon latency meets the tolerated delay. Similarly, another QoE utility is defined based on whether the latency or the framerate requirements is satisfied\cite{9097455}.
% From the general perspective of VR interaction streaming, \cite{9565222} defined the QoE based on the combination of viewpoint prediction, video quality, and video jitter. 
 % This reduces the bandwidth costs and the delay of background content transmission and ensures the real-time performance of interactive content.

\subsection{Motivation and Contributions}
Most related works on DTEC rely on RL algorithms which are only trained offline to make task-offloading decisions to MEC~\cite{9174795,9447819,9170905}.
However, these works ignore the impact of time-varying desynchronization between the physical and digital worlds on the effectiveness of resource allocation strategy. To address this, we propose a continual RL framework that enables continual learning in dynamic environments. The framework allows the DTEC to adapt the strategy using constant DT updates. This mitigates the desynchronization between the physical world and the digital world and ensures optimal allocation strategy. 
Additionally, conventional performance metrics such as latency, energy consumption, transmission rate, and image quality are utilized to measure the system performance in~\cite{9174795,9447819,9170905,7915547,8416431,9789210,9881517,9350227}. These metrics do not guarantee a comprehensive evaluation of MEC-supported VR performance for user-centric applications. Although~\cite{10066518,10101693,10005856,9097455, 9565222} have proposed customized QoE metrics for different scenarios, common-sense QoE metrics are still needed to provide an immersive and interactive experience in VR content streaming applications. Therefore, it is necessary to define customized QoE metrics from different perspectives to evaluate MEC-supported VR performance. 
Since VR applications require real-time rendering of a massive amount of data, which can put a strain on computation and communication resources. 
Attention-based resolution can reduce this strain by rendering the most important parts of the scene in high definition based on users' interaction~\cite{9506358,du2022exploring,du2022attention}. Compared to object attention-based resolution \jiadong{with complex scenes}, gaze attention-based resolution can significantly reduce the computational demands of rendering complex virtual scenes while maintaining high visual fidelity.
% Object attention-based resolution can help address this issue by providing users with a more immersive experience by rendering the most important parts of the scene in high detail based on interaction~\cite{9506358,du2022exploring,du2022attention}. 
% This motivates us to design customized attention-based QoE and consider adaptive resolution strategies for an immersive experience. 

\textit{Since attention-based rendering can reduce both computation and bandwidth costs,
we design our DTEC continual RL framework to consider a customized QoE as a combination of latency and attention-based resolution. This ensures an immersive and interactive Metaverse experience.}

The contribution of this paper can be summarized as follows:
\begin{itemize} 
\item We model a DTEC system and DTs of the users to \jiadong{ assist in making optimal strategies on the MEC system. The user's DT involves the personal FoV attention level information and the QoE. DTEC involves the communication transmission rate, computation capacity of all users and system fairness.}
\item We formulate an optimization problem that maximizes the QoE by combining the latency and gaze attention-based resolution perception for VR users. The problem optimizes the action space toward optimal resource allocation. The problem leverages attention-based resolution of the VR content to allocate CPU resources at the edge server, and bandwidth resources in the edge network.
\item We propose a continual reinforcement learning framework in DTEC to solve the optimization problem. This framework enables constant updates based on real-world execution for optimal strategy in time-varying environments. This mitigates the desynchronization between the physical and digital worlds. We define the CRL's reward function as the sum of QoE and two penalty terms based on QoE and horizon-fairness QoE (hfQoE) constraints.
\item We introduce freshness prioritized experience replay - continual deep deterministic policy gradient (FPER-CDDPG) to further enhance the performance of continual learning in the presence of time-varying DT updates. This approach replaces the original uniform experience replay with freshness prioritized experience replay to enable learning from the experience with the highest priority stored in the replay buffer.
\item We evaluate our proposed framework using extensive experiments. FPER-CDDPG achieves the highest QoE when allocating the MEC resources compared to benchmarks, especially over the long-term run.

\end{itemize}

The rest of this paper is organized as follows. The DTEC system model is introduced and the problem is formulated in Section~\ref{sec:system}. Section~\ref{sec:algo} presents our proposed CRL framework and algorithm FPER-CDDPG. Section~\ref{sec:evaluation} and Section~\ref{sec:numericalresults} present the system evaluation and the numerical results. We conclude our contribution and findings in Section~\ref{sec:conclusion}.

\section{System model}
\label{sec:system}
%Due to the limited computation and battery at head-mounted displays (HMDs), MEC has been a primary pillar for immersive Metaverse. In this paper, the 
% The VR users can Each MEC server, which collocates with the base station (BS) can assist in rendering the different GoP of VR content requested by users thus realising fast delivery of the VR content. As illustrated in Fig.~\ref{algorithm} in the physical world, there are sub-6GHz edge base stations (BS), and $K$ users wearing HMDs. The sub-6GHz BS operates at microwave and it is equipped with an edge server with the capability of providing computation and storage. The HMD can communicate with the BS.

\begin{figure*}[t]
  \centering
  \includegraphics[width=7in]{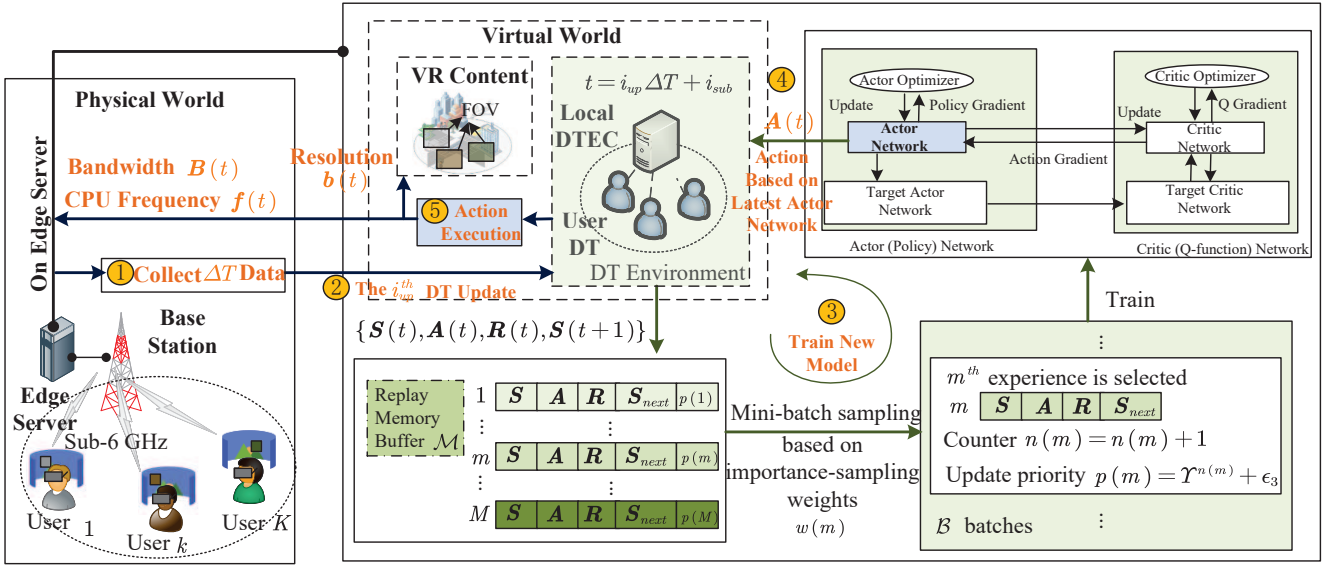}\\
  \caption{Digital twin empowered edge computing and the workflow of the continual RL. (i) Collect Data from the real world, (ii) update the DT in the digital world based on the real data in the physical world, (iii) train the DTEC resource allocation behavioral model based on the updated DT data through the proposed scheme (i.e., FPER-CDDPG), (iv) make the decision action based on the latest trained actor-network, (v) execute the action in both the physical world and virtual world. }
  \label{algorithm}
\end{figure*}

Each MEC server renders different groups of pictures (GoPs) of VR content requested by users. For this purpose, the MEC servers are equipped with sufficient computation and storage resources. This ensures low delivery latency of the VR content. The server collocated with a base station (BS) that relies on sub-6GHz technology %(i.e., microwave) 
for communicating with its users. Fig.~\ref{algorithm} illustrates $K$ users wearing HMDs to communicate with the BS in the physical world. 

%The computing energy consumption can be formulated as $E_{k,a}=\pi F_{k,a}^{3},$ where $\pi$ is the effective capacitance coefficient of the processor chip at the server. The maximum energy consumption constraint of the server is $E_{max}$. 
\subsection{Attention-based VR Content}

\begin{figure}[t]
  \centering
  \includegraphics[width=3.4in]{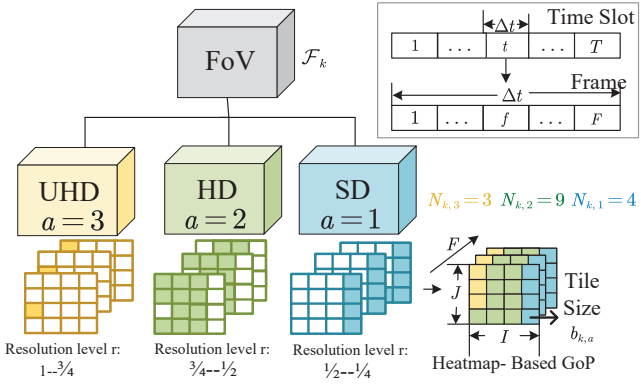}\\
  \caption{Illustration of the attention-based VR video content for the $k^{th}$ user at time slot $t$. \jiadong{With $I=4$, $J=4$, and $F=3$ frames, showing heatmap-based GoP with attention-level-based tile sizes $N_{k,1}=3$, $N_{k,2}=9$, and $N_{k,3}=4$.}}
  \label{FOV}
\end{figure}

As illustrated in Fig.~\ref{FOV}, the 360° VR video content with the spherical feature is projected to the two-dimensional (2D) plane. For the $k^{th}$ user ($k\in \mathcal{K}$) at time slot $t$, the video within the FoV $\mathcal{F}_{k}(t)$ can be uniformly cropped into $I\times J=N$ tiles. 
Inspired by the hierarchical human vision, we define the visual attention level ($a\in \{1,2,3\}$), and assign the tiles to three resolution levels accordingly. The three attention levels correspond to three resolutions when encoding the video frames: $a=3$ ultra-high definition (UHD), $a=2$ high definition (HD), and $a=1$ standard definition (SD). 
At time slot $t$, each attention level has $N_{k,a}$ tiles and remains fixed for a group of frames $F$, referred to as a GoP $G_{k}(t)=\sum_{a}g_{k,a}(t)$. The data size of each tile at attention level $a$ is $g_{k,a}(t)=N_{k,a}\times b_{k,a}(t) \times F$, where $b_{k,a}(t)$ is the size of each tile at attention level $a$ and with resolution level $r_{k,a}(t)$ at time slot $t$. The higher the attention level $a$ and resolution level $r$, the higher size of $b_{k,a}(t)=r_{k,a}(t)\times b_{max}$.

\subsection{User's DT}
The user's DT is the digital replica of the physical user's features. To assist the DTEC's decision-making of resource allocation, 
the user's DT at time slot $t$ is defined as
\begin{align}
    \mathcal{D}_{k}(t)=\{N_{k,a}(t),\text{QoE}_{k}(t)\}, a\in \{1,2,3\}.
\end{align}
given that the \jiadong{tiles number at each attention level} $N_{k,a}(t)$ of the user's FoV 
$\mathcal{F}_{k}(t)$ is well-collected and predicted in the user's DT. 

Peak Signal-to-Noise Ratio (PSNR) $\text{PSNR}_{k}(t) $ is defined by the mean square error (MSE) between the originally rendering FoV $\mathcal{I}(t)$ and the distorted FoV $\Tilde{\mathcal{I}}(t)$ delivered to the user. For the $k^{th}$ user, $\text{PSNR}_{k}(t)$ equals to $\mathcal{I}_{k}(t) - \Tilde{\mathcal{I}}_{k}(t)$. Instead of considering the pixel-based PSNR, a binary function $\text{MSE}_{k}(t)=(\mathcal{I}_{k}(t)-\Tilde{\mathcal{I}}_{k}(t))^2$ is designed, in which $\mathcal{I}_{k}(t)=1$ and $\Tilde{\mathcal{I}}_{k}(t)\in \{0,1\}$. Specifically, $\Tilde{\mathcal{I}}_{k}(t)=0$ denotes the FoV failed to be delivered $(T_{k}> T_{th})$ and $\Tilde{\mathcal{I}}_{k}(t)=1$ denotes the requested FoV is successfully delivered $(T_{k}\leqslant T_{th})$. The PSNR\cite{9411714} of the $k^{th}$ user at time slot $t$ can be defined as 
\begin{align}
    \text{PSNR}_{k}(t)= 10\text{log}_{10}\left(\frac{1+\epsilon_1}{\text{MSE}_{k}(t)+\epsilon_1}\right),
\end{align}
where we set $\epsilon_1=1$ to avoid the infinite value of PSNR cased by $\text{MSE}_{k}(t)=0$. Based on Weber-Fechner Law\cite{5501894}, we introduce a novel concept called attention-based resolutions perception, which assists in quantitatively evaluating the QoE experienced by individual users: 
\begin{align}
    \text{QoE}_k(t)= \text{PSNR}_{k}(t) \sum_{a} \frac{a N_{k,a}}{N} \text{ln}\left(\frac{ b_{k,a}(t) }{b_{th}} \right),
    \label{qoe}
\end{align}
where $b_{th}$ is the smallest tile size.

\subsection{DTEC}
Similar to the definition of the user's DT, DTEC is the digital representation of the edge computing system. We define DTEC at time slot $t$ as:
\begin{align}
\mathcal{D}_{e}(t)=\{R_{k}(t),f_{k}(t),\text{hfQoE}(t)\}, k\in \mathcal{K}.
%    \mathcal{D}_{e}(t)=\{B_k(t),f_k(t),\mathcal{D}_{k}(t),\text{fQoE}(t)\}, k\in \mathcal{K}.
\end{align}
To be specific, this includes the communication transmission rate $R_{k}(t)$ and computation capacity $f_{k}(t)$ of all $K$ users, and the system fairness $\text{hfQoE}(t)$.

\subsubsection{Communication Model}
%all unicast without overlap for simplicity. There are backup bandwidth can be allocated to users(?)
For the sub-6 GHz link between BS and $k^{th}$ HMD, the theoretical transmission rate in DTEC is given as
\begin{align}
R_{k}(t)=B_{k}(t)log_2\left(1+\frac{P_{k}(t)h_{k}(t)\left(r_{k}(t)\right)^{-\alpha}}{I_{k}(t)+\sigma_{k}^{2}}\right),
\end{align}
where $B_{k}(t)$ and $P_{k}(t)$ are the subchannel bandwidth and the transmit power of the BS to $k^{th}$ user at time slot $t$, $h_{k}(t)$ is the Rayleigh channel gain, $r_{k}(t)$ is the distance between BS and $k^{th}$ user, $\alpha$ is the path loss exponent, $I_{k}(t)$ denotes the inter-cell interference, and $\sigma_{k}^{2}$ is the noise power of the sub-6 GHz link\cite{9536410}.

The calibrated communication latency from the edge BS to the $k^{th}$ user over wireless links can be computed as
\begin{align}
T_{k}^{(d)}(t)=\frac{G_{k}(t)}{\omega \left(R_{k}(t)-\varDelta R_{k}(t)\right)},
\end{align}
where $\omega$ is the compression ratio before transmission, $\varDelta R_{k}(t)$ is the estimated rate bias between the theoretical transmission rate in DTEC and the actual transmission rate, retrieved from the feedback in the physical world.

\subsubsection{FoV Rendering Model}
The resource allocation for the $k^{th}$ user computing can be expressed as ${f}_{k}(t)$ which indicates the allocated computing capacity for rendering GoP $G_{k}(t)$ at time slot $t$ in the MEC server. Given $f_{max}$ denoting the server's maximum computing capacity, the calibrated rendering latency of $k^{th}$ user's requested FoV can be computed as
%\begin{align}
%T_{k}^{(r)}=\max\left\{\frac{{{g_{k,a}}c_{a}}}{f_{k,a}}, a\in \mathcal{A}\right\},
%\end{align}
\begin{align}
T_{k}^{(r)}(t)= \frac{\sum_a{{g_{k,a}(t)}c_{a}}}{f_{k}(t)-\varDelta f_{k}(t)},
\end{align}
where $\varDelta f_{k}(t)$ is the estimated CPU frequency bias\cite{9174795} between the estimated CPU frequency in DTEC and the actual CPU frequency, retrieved from the physical world, $c_{a}$ denotes the number of cycles required for processing one bit of input data at attention level $a$. 
The total latency of the $k^{th}$ user to display the requested FoV, which includes both rendering and downlink latency, can be computed as
\begin{align}
    T_{k}(t)=T_{k}^{(d)}(t)+T_{k}^{(r)}(t).
    \label{time_latency}
\end{align}

\subsubsection{Horizon-fair QoE}

Horizon-fair QoE \cite{si2022enabling,7588099} over time horizon $t$ can be computed as
\begin{align}
    \text{hfQoE}(t)=1-\frac{2\sigma}{H(t)-L(t)},
\end{align}
where $L(t)=\min\{\text{QoE}_{k}(t),k\in\mathcal{K}, t\in[0,t]\}$ and $H(t)=\max\{\text{QoE}_{k}(t),k\in\mathcal{K}, t\in[0,t]\}$ are the lowest and highest QoE over the time horizon $t$ among $K$ users, respectively.
$\sigma$ is the standard deviation which can be computed as
\begin{align}
\sigma=\sqrt{\frac{1}{K}\sum_{k=1}^{K}\left(\text{avgQoE}_k(t)-\overline{\text{avgQoE}(t)}\right)^2}
\end{align}
with the average QoE of the $k^{th}$ user over time horizon $t$ computed as
\begin{align}
    \text{avgQoE}_{k}(t)=\frac{1}{t}\sum_{t=1}^{t}\text{QoE}_{k}(t),
\end{align}
and $\overline{\text{avgQoE}(t)}$ as the average of all $K$ users' $\text{avgQoE}_{k}(t)$ at the time slot $t$.
\subsection{Problem Formulation}
The resource allocation behavior in DTEC is formulated to maximize the long-term $\text{QoE}$ for immersive VR experience by jointly optimizing the attention level-based tile resolution $\boldsymbol{b}=\{b_{k,a}(t)\},k\in \mathcal{K}, a\in \{1,2,3\}$, bandwidth $\boldsymbol{B}=\{B_{k}(t),k\in \mathcal{K}\}$, and assigned CPU frequency $\boldsymbol{f}=\{f_{k}(t),k\in \mathcal{K}\}$. The problem can be formulated as
\begin{subequations}
\begin{align}
    (\textbf{P0}) \quad
%&\max_{\mathcal{A}}\quad \sum_{i=t}^{\infty}\gamma^{i-t} \left(\sum_{k=1}^{K}\text{QoE}_{k}(t)\right)
   % \label{a} \\
    &\max_{\boldsymbol{b},\boldsymbol{B},\boldsymbol{f}}\quad \sum_{t=0}^{\infty}\sum_{k=1}^{K}\text{QoE}_{k}(t)
    \label{a} \\
    \textbf{s.t.} \quad 
    &\sum_{k} B_{k}(t)\leqslant B_{max},\quad \forall k\in \mathcal{K}, 
    \label{b}\\
 &\sum_{k} f_{k}(t)\leqslant f_{max},\quad \forall k\in \mathcal{K},
    \label{c}\\
    & \text{QoE}_{k}(t)\geqslant \text{QoE}_{k,th},\quad \forall k\in \mathcal{K},
    \label{d}\\
    & \text{hfQoE}(t)\geqslant \text{hfQoE}_{th},
    \label{e}
\end{align}
\end{subequations}
constraint (\ref{b}) denotes the overall system bandwidth which cannot exceed the total bandwidth $B_{max}$, constraint (\ref{c}) denotes the system CPU frequency cannot exceed the maximum frequency $f_{max}$. The constraint (\ref{d}) is imposed to satisfy the user's QoE, whereas (\ref{e}) satisfies the long-term QoE fairness of the system. To this end,  the constraint (\ref{e}) imposes a lower bound, a threshold denoted as $\text{hfQoE}_{th}$.

\section{Proposed Algorithm}
\label{sec:algo}
Continual reinforcement learning (CRL) is an advanced form of RL that focuses on continual learning and adaptation to changing environments~\cite{khetarpal2022towards}. CRL can be particularly useful in DTEC when the conditions of the physical counterpart system encounter changes over time, such as new operating conditions and system bias (i.e., user behavior, communication, and computation bias). In this section, we implement the CRL framework in DTEC to jointly adapt the attention-based resolutions and allocate the resources over long periods of time.
We also introduce a Freshness Prioritized Experience Replay-based Continual Deep Deterministic Policy Gradient (FPER-CDDPG) framework to account for both old and new experiences for more comprehensive learning. 
\subsection{Problem Transformation Based on Continual Reinforcement Learning}
The state space, action space, and reward are defined under the conventional RL framework as follows:
\begin{itemize}
    \item \textbf{State space}: The state $\boldsymbol{S}(t)$ at each time slot $t$ can be defined as 
    \begin{align}
    \begin{split}
            &\boldsymbol{S}(t)=\{\mathcal{D}_{k}(t-1),\mathcal{D}_{k}(t),\mathcal{D}_{e}(t), T_{k}^{d}(t),T_{k}^{r}(t),\\&T_{k}(t)\}, k\in \mathcal{K}
            \end{split}
    \end{align}
    with the information acquired from the user DTs and DTEC itself. 
    \item \textbf{Action space}: The action vector of the whole system can be formulated as $\boldsymbol{A}(t)=\{\mathcal{A}(t)\}$. We use $\mathcal{A}=\{\boldsymbol{b},\boldsymbol{B},\boldsymbol{f}\}$ to denote the whole solution set. Thus, $\mathcal{A}(t)$ is the action at the $t$ time slot.
    \item \textbf{Reward}: The reward function of the whole system at time slot $t$ can be defined a
\begin{align}
    \begin{split}
    &\boldsymbol{R}(t)=\mathcal{R}(\boldsymbol{S}(t),\boldsymbol{A}(t))\\&=\sum_{k=1}^{K}\text{QoE}_{k}(t)-\varpi_{1}\sum_{k=1}^{K}q_{k}^{\text{QoE}}-\varpi_{2}q^{\text{hfQoE}},
    \label{rewardfunction}
        \end{split}
\end{align}
where $\varpi_{1}$ and $\varpi_{2}$ are the penalty coefficients,
\begin{align}
    q_{k}^{\text{QoE}}=\begin{cases} 0,\quad  \text{QoE}_{k}(t)\geqslant \text{QoE}_{k,th},\quad \forall k\in \mathcal{K}\\
1,\quad \text{otherwise},\\
    \end{cases}
\end{align}
\begin{align}
    q^{\text{hfQoE}}=\begin{cases} 0,\quad  \text{hfQoE}(t)\geqslant \text{hfQoE}_{th}\\
1,\quad \text{otherwise}.\\
    \end{cases}
\end{align}
\end{itemize}

Deep reinforcement learning (DRL) is commonly applied in the context of episodic environments with a stationary distribution. The use of episodic environments in DRL enables us to consider the performance of the policy $\chi$ (i.e., $\chi$ is a mapping from states $\boldsymbol{S}$ to the action $\boldsymbol{A}$) during a single episode. Therefore, to guide the learning process, the episode-based objective is defined as follows:
\begin{align}
    \mathcal{J}_{ep}(\chi)=\mathbb{E}\left[\sum_{\tau=0}^{T-1}\sum_{k=1}^{K}\gamma^{\tau}\boldsymbol{R}(t)|\boldsymbol{S}(t)\right],
    \label{j_ep}
\end{align}
where $\gamma\in [0,1)$ is the discount factor that can determine the weight of the future long-term reward, and $\gamma=0$ indicates that only the current time slot $t$ is considered. 

However, the physical system may exhibit non-stationary behavior, such as changes in user behavior (e.g., $N_{k,a}(t)$), communication environmental conditions (e.g., $\varDelta R_{k}(t)$), operational conditions (e.g., $\varDelta f_{k}(t)$). These changes may impact the system performance and the accuracy of the DT model. To mitigate the desynchronization, the DTEC needs to be updated and retrained on a regular basis. This ensures that the DTEC remains accurate and effective over time in non-stationary environments. This may involve incorporating new data, updating algorithms, or adjusting model parameters to reflect changes in the real-world environment.

CRL addresses the issue of learning in non-stationary contexts as opposed to typical DRL algorithms, which learn from a fixed dataset of episodes. Therefore, in DTEC, each round of 'DT update' can be treated as a single 'episode'.
For the continuing environment, the objective is to learn a policy $\chi$ that maximizes the cumulative discounted reward function with an infinite time horizon. Specifically, we seek to maximize the  objective function over an infinite horizon at each point in time $t$ as
\begin{equation}
\begin{split}
&\mathcal{J}_{con}(\chi)=\mathbb{E}\left[\sum_{\tau=0}^{\infty}\sum_{k=1}^{K}\gamma^{\tau}\boldsymbol{R}(t+\tau)|\boldsymbol{S}(t)\right]=\mathcal{J}_{ep}(\chi)+\\&\mathbb{E}\left[\sum_{\tau=T}^{\infty}\sum_{k=1}^{K}\gamma^{\tau}\boldsymbol{R}(t+\tau)|\boldsymbol{S}(t)\right].
\label{j_con}
\end{split}
\end{equation}
\jiadong{The main difference between eq. (\ref{j_ep}) and eq. (\ref{j_con}) is that the former objective only optimizes over a time horizon $T$ until the current episode terminates rather than until the end of the agent's lifetime.}

\subsection{Continual DDPG with Freshness Prioritized Experience Replay}
\begin{algorithm*}[!t]
\small
\label{algo1}
 \caption{Freshness Prioritized Experience Replay-based Continual Deep Deterministic Policy Gradient (FPER-CDDPG)}
 \hspace*{\algorithmicindent}
\textbf{Initialization:} Actor (policy) network $\chi(\boldsymbol{S}|\boldsymbol{\varPhi}_{\chi})$ with parameters $\boldsymbol{\varPhi}_{\chi}$, critic (Q-function) network $\mathcal{Q}(\boldsymbol{S},\boldsymbol{A}|\boldsymbol{\varPhi}_{\mathcal{Q}})$ with parameters $\boldsymbol{\varPhi}_{\mathcal{Q}}$, target actor network $\chi^{'}$ with parameters $\boldsymbol{\varPhi}_{\chi^{'}}\leftarrow{\boldsymbol{\varPhi}_{\chi}}$, target critic network $\mathcal{Q}^{'}$ with parameters $\boldsymbol{\varPhi}_{\mathcal{Q}^{'}}\leftarrow{\boldsymbol{\varPhi}_{\mathcal{Q}}}$, empty replay buffer memory $\mathcal{M}$. Initialize $L(t)$, $H(t)$, $\text{QoE}_{k,th}$, $\text{hfQoE}_{th}$, $\beta_1$, $\beta_2$, $\mu$.
Initialize observation state $\boldsymbol{S}(1)$. 
 \begin{algorithmic}[1]
\STATEx {\textbf{\%Continual framework -- update the DT in every $\varDelta T$: }}
 \FOR {DT Update $i_{up}=1$ to $\mathcal{I}_{up}=\infty $}
\STATE{Users' DT update (e.g., $N_{k,a}(t),t\in [i_{up} \varDelta T,(i_{up}+1) \varDelta T]$, $k\in \mathcal{K}$).}
\STATE{DTEC update (e.g., $\varDelta R_{k}(t)$ estimated rate bias and
$\varDelta f_{k}(t)$ estimated CPU frequency bias, $t\in [i_{up} \varDelta T,(i_{up}+1) \varDelta T]$).}
 \FOR{$i_{sub}=1$ to $\varDelta T$}
 \STATE $t=i_{up} \varDelta T + i_{sub}$.
 \STATE Observe current normalized system state $\boldsymbol{S}^{(nor)}(t)$ and choose normalized continuous action $\boldsymbol{A}^{(con)}(t)$ by the actor network. 
 \STATE Map the normalized continuous action $\boldsymbol{A}^{(con)}(t)$ into true action space $\boldsymbol{A}(t)$. 
 %  \STATEx{{\textbf{//Enforce the finite queue size constraint:}}}
 %\IF {$Q_{j\rightarrow{[1,K+1]}}(t)>Q_{j\rightarrow{[1,K+1]}}^{th}$}
Execute $\boldsymbol{A}(t)$ in the DT environment.
 \STATE Update $L(t)$ and $H(t)$. Observe the reward $\boldsymbol{R}(t)$ based on eq. (\ref{rewardfunction}).
 
 \STATE Observe new state $\boldsymbol{S}(t)$. 
   \STATEx {\textbf{\%Freshness prioritize experience replay: }}
 \STATE {Normalize new state as $\boldsymbol{S}^{(nor)}(t+1)$. Store $\{\boldsymbol{S}^{(nor)}(t),\boldsymbol{A}^{(con)}(t),\boldsymbol{R}(t),\boldsymbol{S}^{(nor)}(t+1)\}$ in the replay buffer $\mathcal{M}$ at index $m$.}
 \STATE{Reset the counter of the $m^{th}$ buffer $n(m)=0$}.
 \FOR {$m=1$ to $M$}
 \STATE Sample a mini-batch $\mathcal{B}$ transitions $\{\boldsymbol{S}^{(nor)}(t),\boldsymbol{A}^{(con)}(t),\boldsymbol{R}(t),\boldsymbol{S}^{(nor)}(t+1)\}$ from $\mathcal{M}$ based on the weights $w(m)$.
   \STATE Compute TD-error $\delta(m)$ in eq. (\ref{TDerror}).
 \STATE{If the $m^{th}$ experience is selected, then update counter $n(m)\leftarrow n(m)+1$.}
\STATE{Update freshness priority $p(m)$ in eq. (\ref{freshnessPER})}.
\STATE{Calculate the importance-sampling weights $w(m)$ in eq. (\ref{ISweight}).}
\ENDFOR
  \STATE Compute target value based on eq. (\ref{targetvalue}). Update critic $\boldsymbol{\varPhi}_{\mathcal{Q}}$ by minimizing the loss in eq. (\ref{criticloss}) with learning rate $\varrho_{\mathcal{Q}}$.
  \STATE Update actor policy $\boldsymbol{\varPhi}_{\chi}$ using the weighted sampled policy gradient with learning rate $\varrho_{\mathcal{\chi}}$.
\STATE Update target networks with $\boldsymbol{\varPhi}_{\chi^{'}}\leftarrow{\varsigma\boldsymbol{\varPhi}_{\chi}+(1-\varsigma)\boldsymbol{\varPhi}_{\chi^{'}}}$ and $\boldsymbol{\varPhi}_{\mathcal{Q}^{'}}\leftarrow{\varsigma\boldsymbol{\varPhi}_{\mathcal{Q}}+(1-\varsigma)\boldsymbol{\varPhi}_{\mathcal{Q}^{'}}}$.
\ENDFOR
\ENDFOR
\end{algorithmic}
\end{algorithm*}
DDPG is an actor-critic algorithm that can learn continuous control policies in high-dimensional action spaces. This makes it well-suited for solving the proposed problem. In fundamental DDPG, the two different DNNs namely critic network $\mathcal{Q}(\boldsymbol{S},\boldsymbol{A}|\boldsymbol{\varPhi}_{\mathcal{Q}})$ which approximates the Q-function, and actor-network $\chi(\boldsymbol{S}|\boldsymbol{\varPhi}_{\chi})$ which approximates the policy function $\chi$. To be noticed, $\boldsymbol{\varPhi}_{\mathcal{Q}}$ and $\boldsymbol{\varPhi}_{\chi}$ denote the weights of the critic and actor-network DNNs. The edge server will obtain the experience buffer memory to train the critic network $\mathcal{Q}(\boldsymbol{S},\boldsymbol{A}|\boldsymbol{\varPhi}_{\mathcal{Q}})$ and the actor-network $\chi(\boldsymbol{S}|\boldsymbol{\varPhi}_{\chi})$.
\jiadong{Additionally, to improve the algorithm stability during the training stage, DDPG introduces two target networks namely target critic network $\mathcal{Q}^{'}(\boldsymbol{S},\boldsymbol{A}|\boldsymbol{\varPhi}_{\mathcal{Q}^{'}})$ and target actor-network $\chi^{'}(\boldsymbol{S}|\boldsymbol{\varPhi}_{{\chi}^{'}})$. These target networks are created as lagged versions of the primary agent networks, utilizing Polyak averaging\cite{lillicrap2015continuous}. Compared to the primary networks, target networks show slower rate of change which significantly enhances the stability of the DDPG algorithm.}

\jiadong{Continual Reinforcement Learning (CRL) encompasses three main categories of approaches: knowledge retention, leveraging shared structure, and learning to learn. However, a significant challenge within the continual learning framework is catastrophic forgetting. This refers to the agent's tendency to overwrite or forget previously acquired knowledge when updating its behavioral functions, strategies, and other aspects based on new experiences.
Explicating knowledge retention is the category for mitigating catastrophic forgetting~\cite{khetarpal2022towards}. Specifically, there are three types of approaches that lie in this category: latent parameter storage approach, distillation-based approach, and rehearsal-based approach. In the latent parameter storage approach, previous task knowledge is leveraged either as network representations or as priors concerning the extent to which each parameter was used in past tasks. By incorporating this knowledge, the agent can benefit to its new training process.  The distillation-based approach improves the network training process by introducing additional auxiliary targets for the network to match. However, both of these methods involve computationally intensive operations. The rehearsal-based approach is a more straightforward method to address catastrophic forgetting. It involves leveraging experience replay stored in a memory buffer. The experience replay allows the agent to learn from a wider range of experiences, including those that occurred in the past but remain relevant for informing the agent's decision-making in the present and future.}

\jiadong{Conventional experience replay selects the experience uniformly from the memory buffer. However,}
the importance of experiences can change over time as the agent is exposed to a new environment. PER can help the agent to adjust its learning priorities to the current task by giving more weight to experiences that are more relevant to the current task and reducing the weight of experiences that are less relevant. PER modifies the way that the agent samples transition from the replay buffer, which is a memory that stores the experiences of the agent. In conventional RL algorithms, the transitions are sampled randomly from the buffer. However, with PER, the transitions are sampled based on their priority, which is a value that represents how much the transition could improve the current policy. The priority of the experience $j$ is defined based on the TD-error
\begin{align}
    \delta(m)=y-\mathcal{Q}\left(\boldsymbol{S}(t),\boldsymbol{A}(t)|\boldsymbol{\varPhi}_{\mathcal{Q}})\right),
    \label{TDerror}
\end{align}
with $y$ as the target value
\begin{equation}
    y=\boldsymbol{R}(t)+\gamma\mathcal{Q}^{'}\left(\boldsymbol{S}\left(t+1\right),\chi^{'}\left(\boldsymbol{S}(t+1)|\boldsymbol{\varPhi}_{\chi^{'}}\right)|\boldsymbol{\varPhi}_{\mathcal{Q}^{'}}\right),
    \label{targetvalue}
\end{equation}
where $\mathcal{Q}^{'}$ indicates the Q-value calculated based on the target critic network, $\chi^{'}\left(\boldsymbol{S}(t+1)|\boldsymbol{\varPhi}_{\chi^{'}}\right)$ indicates the normalized continuous action instructed by the target actor network ${\chi^{'}}$ given the next normalized state $\boldsymbol{S}(t+1)$.

Since the experiences with large temporal difference (TD)-errors and large negative TD-errors are more likely to be associated with successful attempts and disastrous attempts, respectively. 
The priority of the experience $j$ is
\begin{align}
    p_{m}=\left|\delta(m)\right|+\epsilon_2
    \label{priority}
\end{align}
with $\epsilon_2$ as a small positive constant. Then the probability of the sampling experience $m$ is defined as
\begin{align}
    P(m)=\frac{p_{m}^{\beta_1}}{\sum_{m=1}^{\mathcal{M}}p_{m}^{\beta_1}}
    \label{probabilityofSE}
\end{align}
with $\beta_1$ as the degree of priority. It is a scalar parameter that determines the degree of prioritization of each experience. When $\beta_1$ is set to 0, the algorithm behaves like a standard replay buffer with uniform sampling. As $\beta_1$ increases, the prioritization of the experiences becomes more important, and the algorithm becomes more focused on replaying experiences with high priority. However, too high $\beta_1$ values may lead to overfitting and instability in learning. 

In order to handle the challenge of training process oscillation or divergence, the importance-sampling weights are used in the computation of weight changes
\begin{align}
    w(m)=\left(\mathcal{M} \cdot P(m)\right)^{-\beta_2},
    \label{ISweight}
\end{align}
where $\mathcal{M}$ is the replay buffer size of the buffer memory, and $\beta_2$ is the correction control scalar parameter that controls the level of importance sampling correction when calculating the sampling weights for the replay buffer. Importance sampling is a technique used to adjust the distribution of samples drawn from the replay buffer to match the target distribution used in the learning update. When $\beta_2$ is set to 1, the importance sampling correction is fully applied. When $\beta_2$ is set to 0, the importance sampling correction is not applied, and the sampling weights are equal for all experiences.

For experiences with similar TD errors, there may have some experiences being squeezed out without being replayed. To this extent, the freshness discounted factor $\mu$ which is close but less than 1 is defined\cite{ma2022fresher}. Based on the freshness, the updated priority of the experience $j$ by considering the freshness is defined as
\begin{align}
    p(m)=\mu^{n(m)}\left|\delta(m)\right|+\epsilon_3,
    \label{freshnessPER}
\end{align}
where $n(m)$ counts the number of times that the experience $m$ is replayed, and $\epsilon_3$ is a small positive constant.

During the training process, a mini-batch $\mathcal{B}$ is sampled based on the importance-sampling weights $w(m)$ in eq. (\ref{ISweight}) from the buffer memory to compute the target value $y$. Then, the critic network parameters can be updated based on a learning rate $\varrho_{\mathcal{Q}}$ by minimizing the loss function 
\begin{equation}
\mathcal{L}(\boldsymbol{\varPhi}_{\mathcal{Q}})=\frac{1}{\mathcal{B}} \sum\left(y-\mathcal{Q}(\boldsymbol{S}(t),\boldsymbol{A}(t)|\boldsymbol{\varPhi}_{\mathcal{Q}})\right)^2,
\label{criticloss}
\end{equation}
where $\mathcal{B}$ is the batch size. \jiadong{The training process pseudocode of the proposed algorithm FPER-CDDPG is given in \textbf{Algorithm 1}. }

\begin{table}[t]
\centering
\caption{Summary of simulation parameters}
\begin{tabular}{llll}
\hline
\multicolumn{4}{|c|}{\textbf{System Parameters}}                                                                                                                                     \\ \hline
\multicolumn{1}{|c|}{Parameters}          & \multicolumn{1}{l|}{Value}            & \multicolumn{1}{c|}{Parameters}              & \multicolumn{1}{l|}{Value}                        \\ \hline
\multicolumn{1}{|l|}{$F$}                 & \multicolumn{1}{l|}{$16$}               & \multicolumn{1}{l|}{$I,J,N$}                   & \multicolumn{1}{l|}{$4,4,16$}                          \\ \hline
\multicolumn{1}{|l|}{$K$}                 & \multicolumn{1}{l|}{$4$}                & \multicolumn{1}{l|}{$T_{th}$}                & \multicolumn{1}{l|}{{$150ms$}}               \\ \hline
\multicolumn{1}{|l|}{$\varDelta T$}        & \multicolumn{1}{l|}{$100$}              & \multicolumn{1}{l|}{$P_{k}$}                 & \multicolumn{1}{l|}{$1W$}                           \\ \hline
\multicolumn{1}{|l|}{$B_{max}$}           & \multicolumn{1}{l|}{$10M$}              & \multicolumn{1}{l|}{$f_{max}$}               & \multicolumn{1}{l|}{$15GHz$}                        \\ \hline
\multicolumn{1}{|l|}{$c_{1}$}             & \multicolumn{1}{l|}{$800$ \text{cycles/bit}} & \multicolumn{1}{l|}{$c_{2}$}                 & \multicolumn{1}{l|}{$900$}              \\ \hline
\multicolumn{1}{|l|}{$c_{3}$}             & \multicolumn{1}{l|}{$1000$ \text{cycles/bit}}  & \multicolumn{1}{l|}{$r_{k,1}$}                 & \multicolumn{1}{l|}{{$[1/8,1/4)$} }                 \\ \hline
\multicolumn{1}{|l|}{$r_{k,2}$}             & \multicolumn{1}{l|}{$[1/4,1/2)$}      & \multicolumn{1}{l|}{$r_{k,3}$}                 & \multicolumn{1}{l|}{{$1$}}                    \\ \hline
\multicolumn{1}{|l|}{$b_{max}$}           & \multicolumn{1}{l|}{$12441600$\text{bit}\tablefootnote{This is calculated based on the 4K resolution (i.e., $(3840\times 2160) \text{pixel} \div N \times 8 \text{bit/pixel} \times 3=12441600\text{bit}$.  $3$ represents the Red Green Blue (RGB) colours. ).}}         & \multicolumn{1}{l|}{$b_{th}$}                & \multicolumn{1}{l|}{{$460800$bit}\tablefootnote{This is calculated based on the standard definition (SD) resolution (i.e., $(640\times 480) \text{pixel} \div N \times 8 \text{bit/pixel}   \times 3=460800 \text{bit}$).}}                 \\ \hline
\multicolumn{1}{|c|}{$\text{QoE}_{k,th}$} & \multicolumn{1}{l|}{$9.8645$}                 & \multicolumn{1}{l|}{$\text{hfQoE}_{th}$}  & \multicolumn{1}{l|}{$0.97$}                         \\ \hline
\multicolumn{1}{|c|}{$\epsilon_1,\omega$} & \multicolumn{1}{l|}{$1,300$}                 & \multicolumn{1}{l|}{$\varpi_{1},\varpi_{2}$}  & \multicolumn{1}{l|}{$2,2$}                         \\ \hline
\hline
\hline

\multicolumn{4}{|c|}{\textbf{Algorithms Parameters}}                                                                                                                                  \\ \hline
\multicolumn{1}{|l|}{Parameters}          & \multicolumn{1}{l|}{Value}            & \multicolumn{1}{l|}{Parameters}              & \multicolumn{1}{l|}{Value}                        \\ \hline
\multicolumn{1}{|l|}{$\gamma$}            & \multicolumn{1}{l|}{$0.99$}           & \multicolumn{1}{l|}{$\varrho_{\mathcal{Q}}$} & \multicolumn{1}{l|}{$0.0002$}                       \\ \hline
\multicolumn{1}{|l|}{$L_{ac}$}            & \multicolumn{1}{l|}{$5$}           & \multicolumn{1}{l|}{$n_{ac,1\rightarrow (L_{ac}-1)}$} & \multicolumn{1}{l|}{$256$}                       \\ \hline
\multicolumn{1}{|l|}{$L_{cr}$ }            & \multicolumn{1}{l|}{$5$}           & \multicolumn{1}{l|}{$n_{cr,1\rightarrow (L_{cr}-1)}$} & \multicolumn{1}{l|}{$256$}                       \\ \hline
\multicolumn{1}{|l|}{$\mathcal{M}$}       & \multicolumn{1}{l|}{$10000$}            & \multicolumn{1}{l|}{$\varrho_{\chi}$}        & \multicolumn{1}{l|}{$10^{-7}$} \\ \hline
\multicolumn{1}{|l|}{$\mathcal{B}$}       & \multicolumn{1}{l|}{$64$}               & \multicolumn{1}{l|}{$\varsigma$}             & \multicolumn{1}{l|}{$0.01$}                         \\ \hline
\multicolumn{1}{|l|}{$\beta_1$}           & \multicolumn{1}{l|}{$0.9$}              & \multicolumn{1}{l|}{$\beta_2$}               & \multicolumn{1}{l|}{$0.8$}                          \\ \hline
\multicolumn{1}{|l|}{$\mu$}               & \multicolumn{1}{l|}{$0.95$}             & \multicolumn{1}{l|}{$\epsilon_2,\epsilon_3$}            & \multicolumn{1}{l|}{$0.001,0.001$}                        \\ \hline
\end{tabular}
\label{tab:parameters}
\end{table}

\section{Evaluation}
\label{sec:evaluation}
\ahmad{In this section, we evaluate the performance of our proposed algorithm, FPER-CDDPG, against other benchmarks. After discussing the simulation settings, we introduce the dataset and the preprocessing steps to prepare the data for our experiments. We then report the benchmarks and comparison results.}

\subsection{Simulation Setting}
We assume the edge server is located in the middle of a building ( coordinates $[0m,0m]$) with $K=4$ VR users communicating with it. The locations of the users are $\mathcal{P}_{1}=[23m,1m]$, $\mathcal{P}_{2}=[20m,0m]$, $\mathcal{P}_{3}=[10m,5m]$, $\mathcal{P}_{4}=[15m,5m]$.
The distance-dependent path-loss exponent $\alpha=4$ and the channel noise power is $\sigma^2=-174$dBm. \ahmad{Table \ref{tab:parameters} lists the values of the parameters that are used to simulate the system. The table also specifies the values of essential parameters of the algorithms, including the critic and actor networks}.  Three hidden layers, each with 256 neurons, are included in the actor and critic networks, respectively. Our simulation was performed using Spyder Python 3.7 and run on the NVIDIA GeForce RTX 3060 Laptop GPU.

\begin{table}[!t]
\centering
\caption{Sample pre-processed dataset: Attention-based tile numbers $N_{k,a}$ over $3$ time slots for $4$ VR users.}
\begin{tabular}{|l|l|l|l|l|l|l|l|l|l|} 
\hline
Time Slot                      & \multicolumn{3}{l|}{t=1} & \multicolumn{3}{l|}{t=2} & \multicolumn{3}{l|}{t=3}  \\ 
\hline
\textbf{Attention Level $a$} & 1 & 2  & 3               & 1 & 3  & 3               & 1 & 2  & 3                \\ 
\hline
\hline
\hline
User $k=1$                       & 0 & 13 & 3               & 0 & 13 & 3               & 0 & 15 & 1                \\ 
\hline
User $k=2$                       & 4 & 11 & 1               & 4 & 11 & 1               & 1 & 13 & 2                \\ 
\hline
User $k=3$                       & 7 & 8  & 1               & 7 & 8  & 1               & 7 & 8  & 1                \\ 
\hline
User $k=4$                       & 11& 4 & 1               & 11 & 4 & 1               & 11 & 4 & 1                \\
\hline
\end{tabular}
\label{tab:dataset}
\end{table}
\subsection{Dataset Preparation}
\ahmad{The attention level data\footnote{https://github.com/xuyanyu-shh/VR-EyeTracking} is processed based on the collected eye gaze data from~\cite{8578657}. Specifically, the dataset used in the study consists of 208 HD dynamic 360° videos sourced from YouTube, and the content ranges from indoor scenes, outdoor activities, music shows, sports games, documentation, and short movies. During the user experiment, 45 participants wore HTC VIVE HMD to play the 360° video clips using the Unity game engine. Besides, a '7invensun a-Glass' eye tracker is integrated into the HMD to capture the viewers' gaze. The viewer's heading and gaze direction was recorded, resulting in the eye gaze dataset that we used. The tracker is recalibrated after each group when participants remove the HMD.} 

% The data are collected from the participation of 45 individuals. During the experiment, the Unity game engine was utilized to display the video scenes and record the viewer's heading and gaze direction. The experiments employ an HTC VIVE HMD to play the 360° video clips, and a '7invensun a-Glass' eye tracker is integrated into the HMD to capture the viewers' gaze. The starting point for each video is fixed at 0° latitude and 180° longitude. %To mitigate fatigue while viewing 360° videos, short breaks of 20 seconds are enforced between two video clips, and a longer break of 3 minutes is provided in the middle of each video group. 

In our experiment, we created a long-term dataset for every single user by randomly combining eye gaze coordinate data files from different users and videos, resulting in 8,000,000 frames of data for each user. From this data, we calculated attention-based tile numbers ($N_{k,a}$) based on $F$ frames of gaze centre coordinates. As shown in Fig.~\ref{FOV}, the tile laying the gaze coordinate was assigned an attention level of 3, adjacent tiles that were 2 tiles wide were assigned an attention level of 2, and the remaining tiles were assigned an attention level of 1. The heatmap-based GoP was then created using the assigned attention levels based on the total $F$ frames. 
% Table~\ref{tab:dataset} demonstrates sample data after pre-processing the eye gaze data. The pre-processed data associates  the attention-based tile numbers $N_{k,a}$ in users' DTs over $3$ time slots. In each time slot, the sum of the tile numbers for all $3$ attention levels is $N=16$. covers $3$ time slots and the user's attention changes across the slots. This

\ahmad{Table~\ref{tab:dataset} demonstrates sample data after pre-processing the eye gaze data. The table assigns tiles numbers to attention levels based on the user's attention, i.e., attention-based tiles number $N_{k,a}$, for $4$ users. The table gives an example of the change in users' DTs over $3$ time slots. In each time slot, the sum of the tile numbers for all $3$ attention levels is $N=16$. Notably, for user $k=1$, at time slots $t=1$ and $t=2$, the tile number $N_{1,1}=0$ at attention level $1$, possibly indicating that the user's gaze frequently shifts among the $16$ frames or drops on the center tiles, resulting in higher attention levels for all tiles. Furthermore, users $k=3$ and $k=4$ exhibit stable $N_{k,a}$ values, which could be attributed to fixed gaze patterns within adjacent time slots.
%In each DT update round $i_{up}$, we incorporated the most recent $\varDelta T=100$ time slot real-world data to update the status of the DT, facilitating model training. 
In each DT update round $i_{up}$, The most recent $\varDelta T=100$ time slots of real-world data are used to update the status of the DT, facilitating model training. The overall DT update time horizon is set to $\mathcal{I}_{up}=5000$ rounds for comparison, instead of using an infinite horizon.}
%% Subsection: Benchmarks and Comparison 
\subsection{Benchmarks and Comparison}
\begin{figure}[!t]
  \centering
\includegraphics[width=3.4in]{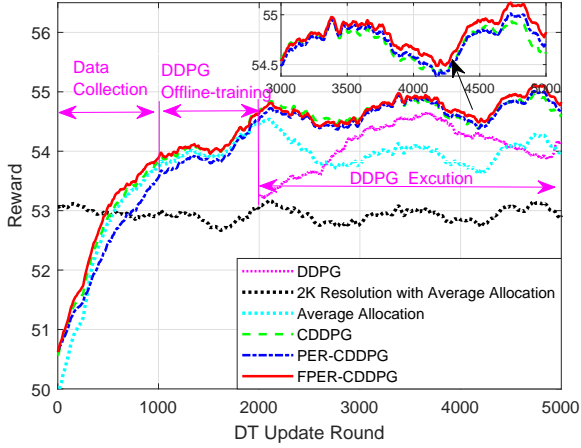}\\
  \caption{The reward performance of different algorithms.}
  \label{allcomparison}
\end{figure}
%\textbf{Benchmarks.} 
To evaluate the performance of our proposed algorithm, we use five benchmarks: 
\begin{itemize}
    \item \textit{\textbf{DDPG:}} offline training of the DDPG algorithm using $1000$ episodes, utilizing the initial $1000$ DT update data. Online execution of the trained DDPG model follows offline training.
    \item \textit{\textbf{2K Resolution with Average Allocation:}} fixed resolution for different attention levels, the communication and the computation resources are uniformly allocated,
    \item \textit{\textbf{Average Allocation:}} adapting attention-based resolution with the communication and the computation resources are uniformly allocated,
    \item \textit{\textbf{CDDPG:}} continual DDPG with DT update constantly,
    \item  \textit{\textbf{PER-CDDPG:}} continual DDPG with prioritized experience replay  with DT update constantly.
\end{itemize}

\begin{figure}[!t]
  \centering
\includegraphics[width=3.4in]{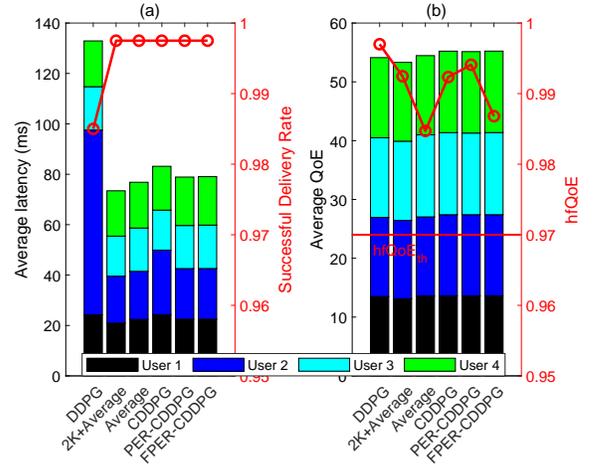}\\
  \caption{Comparison of different algorithms at the latest $100$ time slot (a) average latency and the system successful delivered rate, (b) average QoE and the system horizon-fairness QoE.}
  \label{allcomparison_detail}
\end{figure}

%\textbf{Results.}  
We first compare the reward performance of different benchmarks over DT updates. Fig.~\ref{allcomparison} illustrates that the reward remains at a steady level (i.e., fluctuating around $53$) for the 2K Resolution with Average Resource Allocation benchmark. We attribute that to the lack of attention-based adaptability of the resolution to increase the user's definition perception, as defined in eq.(\ref{qoe}). The Average Allocation benchmark shows a better performance after $500$ DT update rounds, indicating the importance of attention-based adaptive resolution.
After $2000$ DT update rounds, the DDPG execution mostly outperforms the Average Allocation benchmark, showing the effectiveness of the offline-trained DDPG model. However, offline-trained DDPG does not account for the DT updates  coming from the physical world during the online execution.
CDDPG, PER-CDDPG, and FPER-CDDPG outperform DDPG significantly. The continuous RL (CRL) framework uses the latest DT data for model training and online execution, leading to higher performance. Additionally, CDDPG uniformly selects experience data from the replay buffer for training, ignoring the potential influence of some crucial experiences in the long-term DT update. Thus, it shows the fundamental performance as the continual framework. PER-CDDPG slightly outperforms CDDPG due to the selection of experience data based on priority (in eq.(\ref{priority})) and importance-sampling weights (in eq.(\ref{ISweight})) for training. Furthermore, our proposed FPER-CDDPG outperforms all benchmarks after longer rounds of DT updates. The discount factor introduced to the priority in eq.(\ref{freshnessPER}) allows for more comprehensive experience replay for model training, resulting in superior performance.

We then compare comprehensively the benchmarks based on latency, delivery rate, and quality metrics at the latest DT update.
Fig.~\ref{allcomparison_detail} (a) presents the average latency using bars, and the successful delivery rate using red dots (i.e., the ratio of successfully meeting the $T_{th}$ requirement). The offline-trained DDPG without any DT update exhibits the highest latency and, naturally, the lowest successful delivery rate. In contrast to offline-trained DDPG, other benchmarks have the same level of successful delivery rate. The 2K Resolution with Average Resource Allocation benchmark exhibits the lowest latency among all the benchmarks. This is because the 2K resolution provides a relatively small data size of the GoP for rendering. Fig.~\ref{allcomparison_detail} (b) presents the average QoE using bars, and the hfQoE using red dots. The proposed FPER-CDDPG achieves the highest QoE among all benchmarks. However, all benchmarks meet the fairness requirements (i.e., satisfy the $\text{hfQoE}_{th}$ threshold).
\section{Numerical Results}
\label{sec:numericalresults}
% \jiadong{In this section, we provide numerical results of the proposed algorithm FPER-CDDPG.}
\ahmad{In this section, we investigate the performance of the proposed algorithm, FPER-CDDPG, in different settings and provide numerical results of our findings. This includes the impact of algorithm parameters and different systems on the performance, as well as the computation complexity.}
\begin{figure}[!t]
  \centering
\includegraphics[width=3.4in]{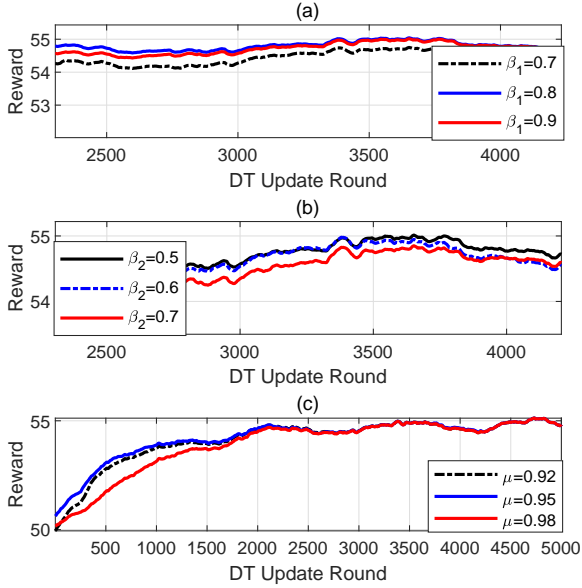}\\
  \caption{The reward performance of different algorithm parameters, (a)$\beta_1$, (b) $\beta_2$, (c)$\mu$.}
  \label{comp2}
\end{figure}
\begin{figure}[t]
  \centering
\includegraphics[width=3.4in]{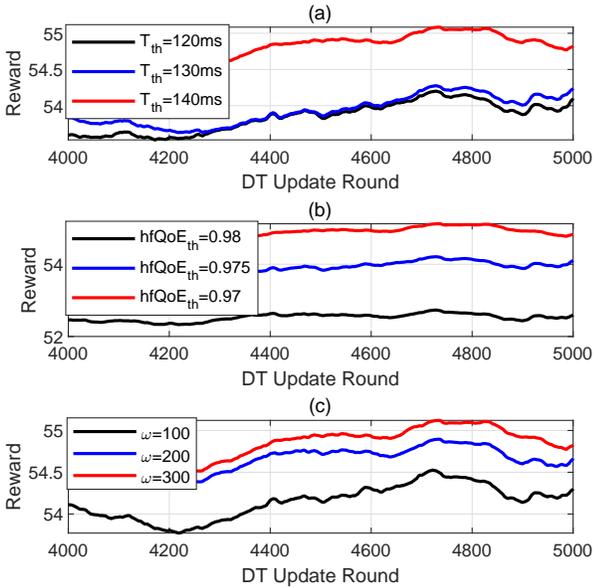}\\
  \caption{The reward of different environment parameters, (a) time threshold $T_{th}$, (b) horizon-fairness QoE $\text{hfQoE}_{th}$, (c) compression ratio $\omega$.}
  \label{comp1}
\end{figure}
\subsection{Impact of Different Parameters}
We first investigate the impact of changing some algorithm parameters on reward performance. Fig.~\ref{comp2} illustrates the impact of changing $\beta_1$, $\beta_2$, and $\mu$ on reward performance. Fig.~\ref{comp2} (a) compares the impact of different degrees of priority $\beta_1$ assigned to each experience. As defined in equation (\ref{probabilityofSE}), a higher value of $\beta_1$ corresponds to a higher probability of replaying that experience. Setting experiences with suitable replay probability leads to more complete learning of the environment before old experiences stored in memory are forgotten.
Fig.~\ref{comp2} (b) shows the influence of the correction scalar $\beta_2$, as defined in equation (\ref{ISweight}). Since relying solely on priority-based importance-sampling can lead to overfitting, $\beta_2$ balances the uniform sampling ($\beta_2=0$) and importance-sampling ($\beta_2=1$) weights.
Fig.~\ref{comp2} (c) investigates the influence of the freshness parameter $\mu$. The difference in performance is not significant when the parameter's value is set close to $1$.

\begin{figure}[t]
  \centering
\includegraphics[width=3.4in]{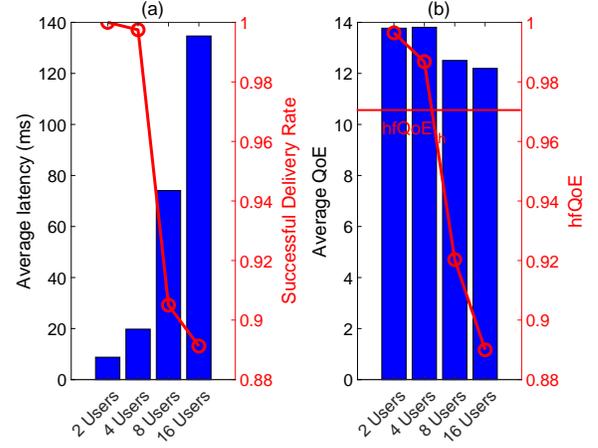}\\
  \caption{System scalability. Increase of (a) average latency and the successful delivery rate and (b) average QoE and the system horizon-fairness QoE of each user as users increase.}
  \label{fig:scalability}
\end{figure}
\begin{figure}[t]
  \centering
\includegraphics[width=3.4in]{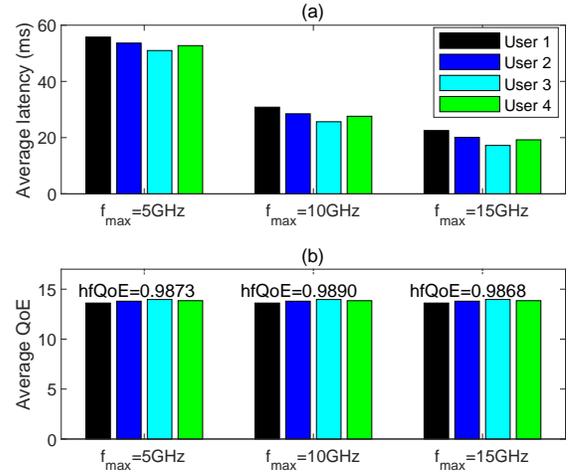}\\
  \caption{System's average latency (a) and QoE (b) for different available CPU frequencies $f_{max}$.}
  \label{cpu}
\end{figure}
We then investigate the impact of changing some environment parameters on reward performance. 
It can be observed in Fig.~\ref{comp1} (a) that when the time threshold $T_{th}$ decreases, it becomes hard for users to meet the requirements, leading to lower rewards. Besides, when the $\text{hfQoE}_{th}$ requirement increases, the reward performance drops, as shown in Fig.\ref{comp1} (b). Moreover, a higher compression ratio $\omega$ results in a faster download of the rendered data from the edge server to the user, leading to higher rewards, as illustrated in Fig.~\ref{comp1} (c).

We examine the system's scalability as the number of users increases while the resources remain the same. Fig.~\ref{fig:scalability} shows the system performance when the total number of users increases exponentially from 2 to 16 while the maximum bandwidth $B_{max}$ and CPU frequency $f_{max}$ remain the same. The system resources become insufficient to accommodate the requirements of each user when the total number of users in the system increases, as indicated by the experimental results in Fig.~\ref{fig:scalability}. The average latency for each user, as depicted in Fig.~\ref{fig:scalability} (a), increases as the number of users increases, which leads to a lower successful delivery rate. More users lower the average QoE of each user, as seen in Fig.~\ref{fig:scalability} (b), which makes the hfQoE threshold harder to meet. One potential solution to overcome this issue is to increase the penalty weight $\varpi_2$, which may help the system meet the hfQoE threshold.

\begin{figure}[t]
  \centering
\includegraphics[width=3.4in]{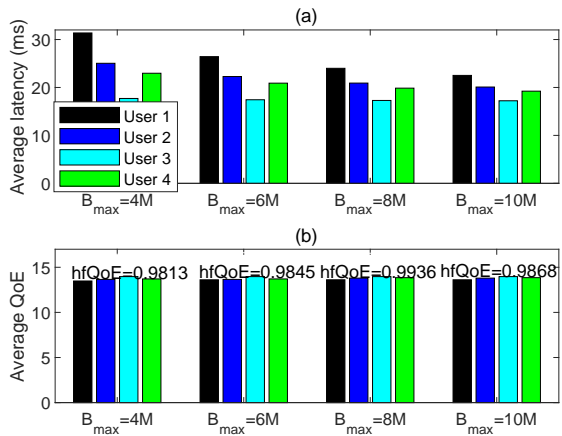}\\
  \caption{System's average latency (a) and QoE (b) for different settings of the available bandwidths $B_{max}$.}
  \label{bw}
\end{figure}

We further examine the system performance as the resources of the edge computing system change. Fig.~\ref{cpu} reveals the average latency and QoE for each user under specific CPU frequencies, while Fig.~\ref{bw} illustrates them under specific bandwidth capacities. Higher CPU frequency increases the computation ability for rendering improves, resulting in a reduction of the average latency of each user (see Fig.~\ref{cpu} (a)). The resolution perception, as given by Eq.(\ref{qoe}), is the main factor for the QoE, since the time threshold is mostly satisfied by all users. The attention-based resolution algorithm we propose maintains a similar QoE level for each user, meeting the hfQoE requirements (see Fig.~\ref{cpu} (b)).
Similarly, as seen in Fig.~\ref{bw} (a), more communication bandwidth increases the communication throughput for each user to receive the rendered content from the edge server, which lowers the average latency of each user. This also enables high QoE and hfQoE to be maintained, as shown in Fig.~\ref{bw} (b).

\subsection{Complexity}
The time complexity for individual critic-actor policy training in CDDPG at DTEC is primarily determined by $\mathcal{O}\left(\sum_{l=0}^{L_{ac}-1}n_{ac,l}n_{ac,l+1}+\sum_{l^{'}=0}^{L_{cr}-1}n_{cr,l^{'}}n_{cr,l^{'}+1}\right)$, where $L_{ac}$ and $L_{cr}$ are the numbers of layers in the actor and critic networks, respectively, and $n_{ac,l}$ and $n_{cr,l^{'}}$ are the numbers of neural nodes in each layer of the actor and critic networks. For PER-CDDPG, where the replay buffer size is denoted as $\mathcal{M}$, the computational complexity must also include the updating and sampling operations, which are $\mathcal{O}(\log_2\mathcal{M})$. In FPER-CDDPG, the only difference is the additional freshness update based on a counter during the updating operation when selecting batch experiences. Thus, the complexity for experience replay is $\mathcal{O}(\mathcal{B}+\log_2\mathcal{M})$, where $\mathcal{B}$ is the batch size.

\begin{table}[t]
\caption{Operation time of different algorithms at DTEC}
\centering
\begin{tabular}{|l|l|l|l|} 
\hline
\textbf{Algorithms} & \textbf{CDDPG}   & \textbf{PER-CDDPG} & \textbf{FPER-CDDPG}  \\ 
\hline
Operation Time (ms) & 0.96             & 6.61               & 10.28                \\ 

\hline
\hline
\hline
\textbf{FPER-CDDPG} & \textbf{2 Users} & \textbf{8 Users}   & \textbf{16 Users}    \\ 
\hline
Operation Time (ms) & 10.05            & 11.57              & 12.21                \\
\hline
\end{tabular}

\label{tab:operationTime}
\end{table}

The computation time for every behavioural model update (i.e., optimal strategy) at DTEC is presented in Table~\ref{tab:operationTime}. The operation time increases significantly in PER-CDDPG and FPER-CDDPG due to the introduction of the experience replay update process, 6.61 ms and 10.28, respectively, compared to CDDPG. Although the number of users increases exponentially (2,8,16 users), the operation time increases slightly (10, 11.5, and 12.2 ms, respectively) indicating that FPER-CDDPG scales up with the number of users. We attribute this increase in operation time to the increase in the size of the action and state space, resulting in an increase in the number of neural nodes at the output layer.

\section{Conclusion}
\label{sec:conclusion}
This paper presented our framework to address the challenge of time-varying desynchronization between the physical and digital worlds in dynamic resource allocation for MEC-enabled VR content streaming. We first defined a customized QoE metric as the combination of latency and gaze attention-based resolution to evaluate the system's performance. By formulating an attention-based QoE maximization problem, we optimized the allocation of CPU, bandwidth resources, and attention-based resolution to enhance the QoE. 
The proposed framework enables continual learning in dynamic environments, which allows the DTEC to adapt its strategy using constant DT updates and ensures optimal allocation strategy. 
The algorithm utilizes the reward function, composed of QoE and penalty terms based on the QoE and hfQoE constraints, to guide the learning process.
We also introduced freshness and the priority of the experience replay to the continual learning, which resulted in our ultimate algorithm (named FPER-CDDPG). This was intended to enhance the continual learning performance in the presence of time-varying DT updates. We conducted extensive experiments using a simulation. Our finding revealed that continual learning in DTEC improved the system performance, compared to that without DT update. The system performance further improved when using freshness-prioritized experience replay in the long-term run. Besides, this paper highlighted the limitations of existing approaches that ignore desynchronization and provided a comprehensive solution as a DT-empowered MEC system. 

In short, incorporating continual learning with attention-based QoE metrics and freshness prioritized experience replay empowers the DTEC to \jiadong{make optimal strategies on both adaptive resolution and resource allocation}
%optimize resource allocation 
and ensure a high-quality user experience.

%\ahmad{Shall we add some future directions?}

\bibliographystyle{IEEEtran}
\bibliography{IEEEabrv,mybib}
\end{document}